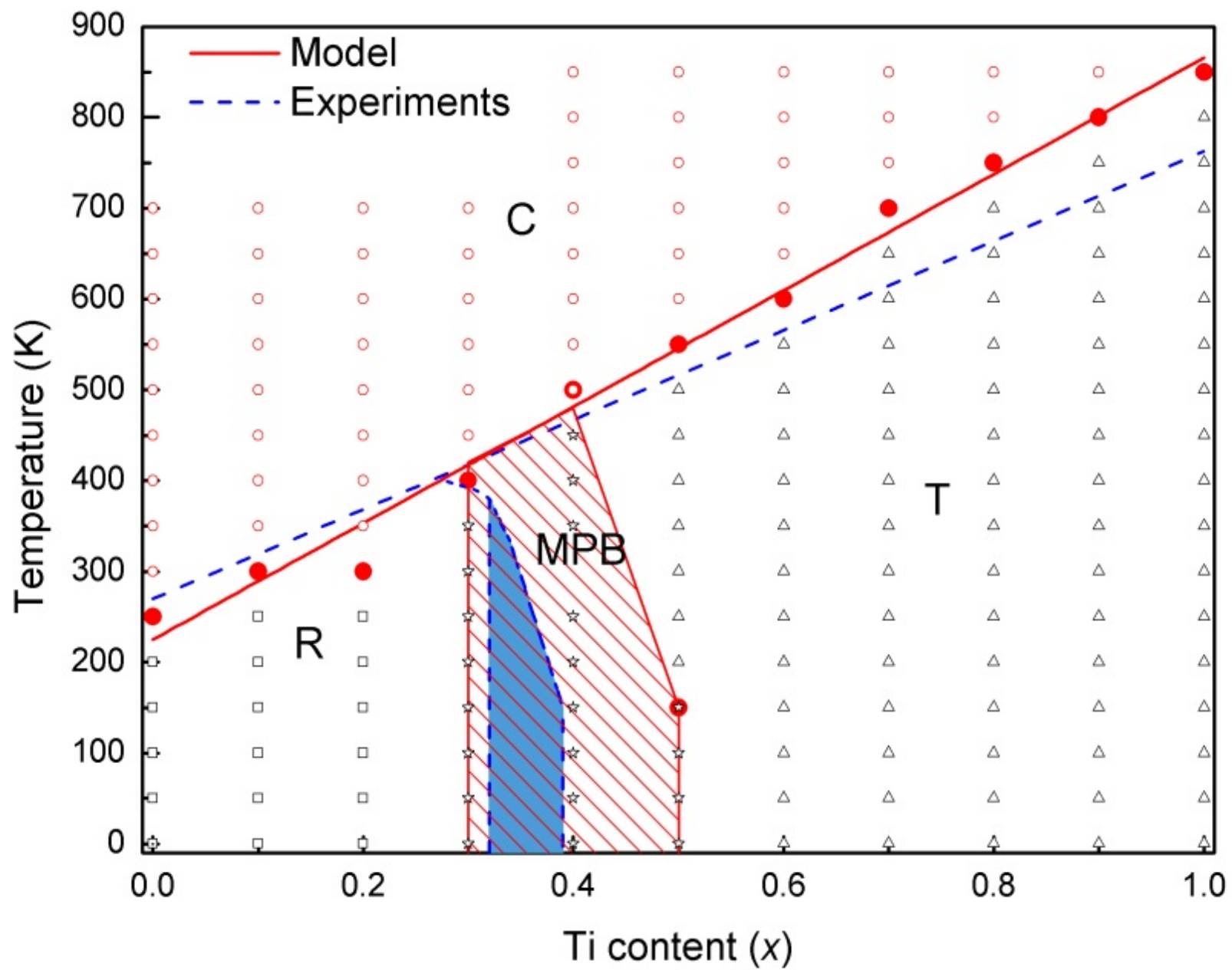

# First principles based atomistic modeling of phase stability in PMN-*x*PT

**M Sepliarsky[1] and R E Cohen[2]**

[1] Instituto de Física Rosario, Universidad Nacional de Rosario – CONICET. 27 de Febrero 210 bis, Rosario, Argentina.

[2] Geophysical Laboratory, Carnegie Institution of Washington, 5251 Broad Branch Road, NW, Washington DC 20015, U. S. A

E-mail: sepli@ifir-conicet.gov.ar

**Abstract**. We have performed molecular dynamics simulations using a shell model potential developed by fitting first principles results to describe the behavior of the relaxor-ferroelectric *(1-x)*$PbMg_{1/3}Nb_{2/3}O_3$-$x$$PbTiO_3$ (PMN-*x*PT) as function of concentration and temperature, using site occupancies within the random site model. In our simulations, PMN is cubic at all temperatures and behaves as a polar glass. As a small amount of Ti is added, a weak polar state develops, but structural disorder dominates, and the symmetry is rhombohedral. As more Ti is added the ground state is clearly polar and the system is ferroelectric, but with easy rotation of the polarization direction. In the high Ti content region, the solid solution adopts ferroelectric behavior similar to PT, with tetragonal symmetry. The ground state sequence with increasing Ti content is R-$M_B$-O-$M_C$-T. The high temperature phase is cubic at all compositions. Our simulations give the slope of the morphotropic phase boundaries, crucial for high temperature applications. We find that the phase diagram PMN-*x*PT can be understood within the random site model.






# 1. Introduction

Relaxor ferroelectric have been give much attention because their extraordinary dielectric and electromechanical properties. For instance, piezoelectric coefficients one order of magnitude larger than in $PbTi_xZr_{1-x}O_3$ (PZT) were observed in single crystals of *(1-x)*$PbMg_{1/3}Nb_{2/3}O_3$-*x*$PbTiO_3$ (PMN-*x*PT) and*(1-x)*$PbZn_{1/3}Nb_{2/3}O_3$-*x*$PbTiO_{1/3}$ (PZN-*x*PT) [1]. Understanding the origin of their behavior is a fascinating area of research, and much progress has been made from extensive experimental [2-5]and theoretical studies[6-9]. Nevertheless many aspects of their rich behavior are not clearly understood due to their complexity. The behavior depends on the interplay of many effects, including composition, chemical inhomogeneities or structural disorder, and involves phenomena in a wide range of lengths and timescales.

PMN-*x*PT is a solid solution of the relaxor PMN and the ferroelectric PT. It has the perovskite structure $ABO_3$ with lead at the A-site, and $Mg^{+2}$, $Nb^{+5}$ and $Ti^{+4}$ occupy B-sites. PMN is a prototypical relaxor crystal with a broad-frequency dependant peak in the dielectric constant [10]. Relaxor behavior is associated with the formation, growing, and freezing of disordered polar nanoregions (PNRs) (see Ref. [11,12]for reviews). Although relaxors do not exhibit long-range order, PNRs may be locally ordered. [13] The existence of PNRs in relaxors is a basic assumption of several models describing relaxor behavior [14-16]. Recent results, however, show that some essential features of the dielectric response in relaxors can be driven by local phenomena without the requirement of PNRs [17-20].

As composition is varied and PT is added, a transition to a ferroelectric phase occurs for concentrations as low as 5% [3], although some characteristics of relaxors such as dispersion, that is frequency dependence, of properties persist in this composition range [21-23]. The underlying origin of the relaxor-ferroelectric crossover with composition is a controversial issue which needs clarification. The proposed mechanisms involve the reorientation and/or growing of PNRs. This low Ti region with rhombohedral symmetry remains up to *x*=0.3, at which a transition region called the morphotropic phase boundary (MPB) begins. The MPB is of particular interest since the largest electromechanical response takes place for concentrations around it [1]. The behavior of the region has been understood in terms of the easy rotation of the polarization directions between the rhombohedral and tetragonal structure [6], and it is not necessarily related to PNR or to complex microstructure. Theoretical and experimental studies in PT under pressure show that the MPB originates due to the presence of different phases with very close free energies, where shallow local minima are embedded in a broad global minimum, and even a pure compound can display a MPB [7]. For example, a MPB is observed in pure $PbTiO_3$ under pressure, and the MPB in solid solutions with PT is due the compositional tuning of the high pressure MPB in PT to ambient pressure [7]. As it is seen in ferroelectrics such as PZT without relaxor behavior or PNRs, the MPB can be induced just by composition, and the polarization can easily rotate between the rhombohedral and tetragonal directions mediated by lower symmetry monoclinic or orthorhombic phases[6].For concentrations above *x*=0.37 the solid solutions are in the tetragonal phase below $T_c$. Here the behavior is similar to PT, although the possibility of an additional low temperature phase has been observed [24].

Relaxor ferroelectrics can be explored at different length scales and levels of complexity, ranging from first principles methods to phenomenological theories. A deep understanding of the behavior of complex solid solution will require linking these different types of studies in a consistent multiscale description. The combination of first-principles calculations with classical atomic models can contribute to investigate systems where a large number of atoms are involved, to study compositional effects in solid solutions, and to compute temperature effects [25]. Model descriptions based on interatomic potentials account naturally for the presence of many factors such as chemical order or compositional heterogeneity. Among these methods, molecular dynamics simulations with shell mod-



els fitted to first-principles calculations revealed powerful to predict the qualitative behavior of pure compounds and solid solutions [26,27].

In the present work we apply a classical shell model potential to study the evolution of PMN-$x$PT as function of concentration and temperature. This is a first-principles model, fit only to first-principles results [28]. Previously this model was used to predict the elastic constants in PMN [29]. Here, we show that the model is able to describe the solid solution in the whole range of concentrations in good agreement with the experimental evidence. According to our results, the different phases observed across the composition results from competing behavior between the end members. While PMN has an average cubic symmetry structure due to the randomness of the atomic displacements from the ideal positions, the presence of PT in the composition favors local order and the development of a polar state. Our model contrasts with the very successful bond-order potential model [30] in that the ionic polarizability is included directly, so that transverse effective charges, and how they change with structure, are included in the model.

## 2. Model and computational details

The shell model has been extensively used in atomistic simulations of oxides because it is a simple method to include the deformation of the electronic structure of an ion (the ionic polarizability) due to the interactions with other atoms. In the shell model, each atom is described as two charged and coupled particles: a higher mass core and a lower mass shell. The model also includes electrostatic interactions among cores and shells of different atoms, and short-range interactions between shells. In this work we use a shell model where core and shell are linked through an anharmonic spring, $V(w)=1/2\, k_2\, w^2+1/24\, k_4\, w^4$, where $w$ is the core-shell displacement. There is an additional penalty term $D(w - w_0)^2$ if $w \geq w_0$, where $w_0 = 0.2$ Å and $D = 10000$ eV, to the core-shell coupling to prevent the shell from drifting off the core, and ensure the potential stability. For the short range interactions between the A-O, B-O and O-O pairs we chose a Rydberg potential, $V(r)=(A+Br)e^{-r/\rho}$, with a shifted-force correction and a cutoff radius of 10 Å. The input data to adjust the parameters were from first principles results within the local density approximation (LDA) [31]. The model parameters were determined by simultaneous least-squares fitting of the end members PMN and PT [28], under the assumption of the transferability of interatomic potentials. In this way, it is possible to study the solid solution in the whole range of concentrations without having to add any extra parameter dependent on concentration or atomic ordering. Model parameters are shown in Table 1.

We used the potential to determine relaxed structures and finite temperature properties of PMN-$x$PT as functions of composition. We used the program DL-POLY [32] within the constant (N, σ, T) ensemble at intervals of 10% in concentration. The runs were performed in system sizes of 12x12x12 unit cells (8640 atoms) with periodic boundary conditions with relaxation times of 0.25 ps and 0.35 ps for the thermostat and barostat respectively. A fraction of the atomic mass is assigned to the shell to permit a dynamical description of the adiabatic condition [33]. A mass of 7 amu is assigned to the Pb, Ti, and Nb shells and 2 amu to the O and Mg shells. The relaxed structures were determined as zero-temperature limit MD simulations. In order to avoid high-energy metastable states successive heating and quenching were performed until forces on individual ions are lower than 0.01 eV/Å. Our relaxed configurations are not necessarily the ground state of the system; in particular chemical diffusion did not occur over the time scale of the simulations, so that chemical ordering or clustering is not variable, but depends on the initial configuration. This corresponds to the experimental situation where chemical ordering or segregation is fixed at the high temperature synthesis conditions. Our MD runs consisted of at least 60000 time steps, with data collect after 20000 time steps, with a time step of 0.4fs, giving run times of 24 ps.



We define the local polarization as the dipole moment per unit volume of a perovskite cell centered at the B-site, and delimited by the 8 Pb near neighbors at the corners of the box. In the calculations we consider contributions from all atoms in the conventional cell, and atomic positions with respect to this center:

$$\mathbf{p} = \frac{1}{v}\sum_i \frac{1}{\omega_i} z_i (\mathbf{r}_i - \mathbf{r}_B)$$

Where $v$ is the volume of the cell, $z_i$ and $\mathbf{r}_i$ denotes the charge and the position of the $i$ particle respectively, and $\omega_i$ is a weight factor equal to the number of cells that the particle belongs. The reference position, $\mathbf{r}_B$, corresponds to the core of the B atom, and the sum extends over 29 particles, including cores and shells of the surrounding ions (8 Pbs and 6 Os) and the shell of the B atom. Note that $\mathbf{p}$ in Eq. 1 is independent of the origin for $\mathbf{r}_i$ vectors, and it is null when atoms are at the ideal cubic positions.

One central point in the simulation is the treatment of the order at the perovskite B-site, which can be occupied by $Nb^{+5}$, $Mg^{+2}$ or $Ti^{+4}$. Chemical order (or disorder) is essential in relaxor behavior, and the current model that fits the existing experimental data in PMN is the presence of chemical ordered regions (CORs) immersed in a disordered matrix [34-36]. The CORs are described by the "random site model" (RSM) [37] in which B-cations display a rock salt type of order, with one sublattice occupied completely by $Nb^{+5}$, and the other contains a random distribution of $Mg^{+2}$ and $Nb^{+5}$ in a 2:1 ratio. We examine here configurations of PMN composed entirely of the RSM model. Starting from PMN, a sequence of configurations with higher Ti content (up to $x$=1) were obtaining by random substitution of $Nb^{+5}$ and $Mg^{+2}$ by $Ti^{+4}$, preserving the neutrality of the system. We performed simulations for three of such sequences (termed 'layouts' below) with different initial PMN configurations in order to analyze the influence of local order on the properties.

## 3. Results

### 3.1. Macroscopic Properties.

Figure 1 shows the dependence on composition of the zero-temperature volume of PMN-$x$PT obtained with the model. The resulting values do not depend on the particular B-cation assignment, so we show just one point at each concentration. We observe that the volume decreases essentially linearly with $x$ in agreement with experimental data (open symbols)[2,3,24,38-40], although the values obtained with the model are slightly lower than the experimental ones. In the model, the volume reduces from 64.10 $Å^3$ in PMN to 61.36 $Å^3$ in PT whereas experimentally the change is from 66.3 $Å^3$[38] to 62.6 $Å^3$ [40] respectively. On the other side, model results are closer to LDA values (open triangles). In LDA the volume of PMN varies between 63.9 $Å^3$ and 64.4 $Å^3$ depending on the symmetry of the structure [9,41], whereas for PT is 60.37 $Å^3$ [42]. Since the model was fitted to reproduce LDA results, we can attribute the underestimation of the experimental volume with the interatomic potential to the difficulty LDA has predicting equilibrium volume in ferroelectrics. In the case of PT, the model gives a larger value than LDA, and this difference fortuitously makes the volume of the solid solution closer to experiments for higher concentrations. Since ferroelectric behavior is very sensitive to volume, we observe that this underestimation is translated to other properties.

Figure 2 shows the evolution of lattice parameters (a) and components of polarization along pseudo-cubic axes (b) as function of the concentration at T = 0 K. Each symbol describes a different layout and solid lines connect the average values at each composition. The symmetry of each configuration is asigned considering an uncertainty of 0.01 Å on the lattice parameters. This uncertainty is



attributed mainly to size effects, and simulations with a larger number of atoms will probably help to reduce it.

Starting at $x = 0$, the simulations indicate that pure PMN is macroscopically cubic and has a negligible polarization. The results are in agreement with experimental observations [38]. First-principles calculations for ordered PMN show polar grouund states [9,41]. Similar results are obtained with the shell model, but we find that the macroscopic polar state vanishes for the disordered system within the RSM. As Ti is introduced in the composition, the cell distorts and the polarization grows. The solid solution at $x=0.1$ and $x = 0.2$ have an average rhombohedral structure with a relatively small net polarization along the [111] direction. A noticeable change in the behavior take place at $x=0.3$, where there is a sharp increment in the magnitude of *P*. From this composition the solid solution definitively becomes ferroelectric and a transition zone begins that separates the region of rhombohedral symmetry from that of tetragonal symmetry. In this compositionally induced MPB, structures with different symmetries as function of the layout and concentration are present. As we observe in Table 2, the estimated phases of the configurations in the region are rhombohedral R (***P***= [*zzz*]), monoclinic $M_B$ (***P***= [*xzz*]), orthorhombic O (***P***= [*0zz*]), monoclinic $M_C$ (***P***= [*0yz*]), tetragonal T (***P***= [*00z*]) and triclinic, Tr (***P***= [*xyz*]). The simulated behavior is consistent with an easy rotation of the polarization direction in the region. The magnitude of ***P*** does not depend on the layout and shows only a small dependence with composition. It changes from approximately 44 $\mu C/cm^2$ at $x=0.3$ to 48 $\mu C/cm^2$ at $x=0.5$. Finally, from $x=0.6$ the behavior is independent of the layouts, and the structure becomes tetragonal with ***P*** along the [001] direction. Both tetragonal distortion and *P* grow as Ti content increase, reaching their maximum values at the ending composition. The values for PT, $c/a = 1.08$ and $P = 66$ $\mu C/cm^2$, are in reasonable agreement with LDA results ($c/a = 1.046$ and $P = 66.7$ $\mu C/cm^2$) and with experiments ($c/a= 1.07$ and $P = 75$ $\mu C/cm^2$).

The above description demonstrates that the model well reproduces the sequence of phases experimentally observed in PMN-*x*PT as *x* is varied. According to this, the behavior can be divided in four different regions, namely: pure PMN, low Ti, medium Ti, and high Ti content. Before we turn to a deeper characterization of each one of them, it is interesting to obtain a global picture of the changes that take place at atomic level as function of the composition. To this purpose figure 3 shows respectively the pair distribution function, PDF, and the local polarization distribution, *W(**p**)*, at four particular concentrations. We observe in the figure how the broad peaks that characterize the PDF and *W(**p**)* of PMN narrow as Ti is incorporated in the solid solution. Since broad distributions are related to the randomness in atomic displacements while sharper peaks reflect more ordered atomic displacements, we conclude that the simulated PMN-*x*PT evolves microscopically from a disordered state to an ordered one with *x*.

*3.2. PMN*

Simulated PMN remains cubic paraelectric at all temperatures in coincidence with experimental observations [38], and hence the nature of its behavior is related to effects that take place at microscopic level. The PDF (*x*=0 in figure 3 (a)) obtained at zero temperature indicates that ions are displaced in different directions from the high-symmetry perovskite positions. Since thermal effects are not present, the width of the peaks reflects the range of distances the pairs can have and the randomness in atomic displacements. The cubic structure in PMN results from the average of uncorrelated displacements of ions from the ideal high symmetry positions. Model results are in general good agreement with EXAFS analysis [43] and LDA calculations [44], although bond distances in present simulations are slightly shorter due to the smaller volume. From the different pair contributions to the PDF we determined that the peak at approximately 2 Å corresponds to the superposition of Nb-O and



Mg-O bonds, and the apparent splitting of ≈ 0.1 Å (shoulder at right side) reflects a larger repulsion of the Mg-O pair with respect to the Nb-O one. The peak at 2.83 Å corresponds to Pb-O and O-O pairs, and shoulders at both sides are from the splitting of distances of the Pb-O bonds. A large Pb off-centering provides a strong contribution to local polarization, and it is commonly observed in relaxors and ferroelectric perovskites [43,45,46]. Here, we estimate an average Pb displacement with respect of the center of the $O_{12}$ cage of 0.29 Å and a maximum displacement of 0.5 Å, which is good agreement with respect to experimental value of 0.4 Å [45]. The peak at 3.46 Å corresponds to Pb-Mg/Nb pairs, where the bond distance of Pb-Nb pairs is slightly larger than for Pb-Mg due to their more repulsive character in agreement with DFT results [19]. Finally, the peaks at 4 Å and 5.65 Å indicate that the overall cubic symmetry is maintained despite the variety of atomic displacements. The randomness in atomic displacements produces individual cells polarized in a wide range of values and directions. Local polarizations are isotropically distributed, and each component of *W(p)* displays a broad distribution centered at zero as it is shown in figure 3. The resulting average cell polarization is $<p>= 25$ μC/cm$^2$, with a maximum $p_{max} = 40$ μC/cm$^2$. These values are underestimated with respect to the LDA ones ($<p>= 38$ μC/cm$^2$ and $p_{max} = 67$ μC/cm$^2$), although values in the last case correspond to ordered structures at the experimental volume [19]; the difference can be partially attributed to the smaller volume considered in the simulations.

The polar glass behavior of PMN is confirmed through finite temperature simulations. Figure 4 displays the temperature dependence of *W(p)* obtaining from the simulations. Due to the cubic symmetry of PMN, the three components of *W(p)* are equivalent, and only one is shown. The inset shows the dependence of the Edwards-Anderson order parameter, $q^{EA}$, given by:

$$\mathbf{q}^{EA} = \frac{1}{N}\sum_{i=1}^{N}\langle \mathbf{p}_i \rangle^2$$

Where the sum extends to all *i* cells in the simulation box. $q^{EA}$ characterizes the polar glass behavior of the system and describes the freezing of its local polarizations. We observe that the distribution narrows with increasing temperature, and for temperatures above 300 K the width remains nearly constant. The Edwards-Anderson order parameter $q^{EA}$ decreases nearly linearly with increasing temperature up to 250 K, and approaches a constant value when the temperature further increases. The simulated temperature behavior of PMN reproduces experimental NMR determinations and is in qualitative agreement with predictions in the framework of the spherical random-bond-random-field (SRBRF) model [47,48]. There is, however, a basic difference between both descriptions. While local polarizations represent polar clusters in the SRBRF model, they describe unit-cell dipoles in the present description. Our results support the random-bond glass behavior of PMN and the validity of SRBRF model description since whether the reorientable polar units represent PNRs or unit-cells is not relevant. The original assumption was made in the light of the accepted picture of relaxors, but our simulations leads to a reinterpretation of meaning of the local parameter. We did not include PNRs in our simulations, and there is no evidence of the presence of them in any of the simulated PMN configurations, neither in PDF nor in *W(p)*. Instead, we show that unit-cell polarizations are enough to capture some of the relevant features of PMN behavior (like the glass transition). According to this, $q^{EA}$ has two contributions. One depends on the temperature and describes the motion of reorientable local polarization with a freezing temperature at $T_f \approx 300$ K. The temperature independent contribution, on the other side, indicates that individual cells remains always polarized. Such "pinned" polarization arises from the different local chemical environment of the cells due to B-cation distribution. As consequence, the average atomic positions are shifted from the ideal perovskite sites even at high temperatures. We note that the pinned local polarizations observed here represent the pinned nanodomains suggested in Ref [48]. Experimentally, two characteristics temperatures characterize the relaxor behavior. The freezing temperature $T_f$ at which the low temperature glassy or nonergodic state transforms to an ergodic state and the Burns temperature $T_B$[13]. While present simulations correctly reproduce the first one, we do not address the appearance of the latter. At $T_B$ PMN transforms from relaxor to a classical paraelectric. The crossover is related to changes in the dynamical behavior of the system.

*3.3. Low Ti*

The solid solution at low Ti concentrations behaves as an inhomogeneous ferroelectric, where glassy and ferroelectric characteristics coexist. The addition of a small amount of Ti in PMN does not produce any noticeable change in the PDF, and the broad distribution of *W(p)* indicates that polar disorder still dominating the behavior of the system as we can observe for *x*=0.2 in figure 3. The asymmetric peak in *W(p)*, which microscopically describes the incipient macroscopic polarization observed in the region, indicates that local polarizations are not completely random. Our simulations reproduce the experimentally observed low Ti region of PMN-*x*PT where relaxor and ferroelectric properties coexists [3,39,49].

We find that the order starts locally: Figure 5 (a) shows the individual pair contributions to the shortest B-O pair, which shows a two-peak structure for the Ti-O bond in contrast to the unimodal distribution for the Nb-O and the Mg-O pairs. The splitting is not directly observable in the PDF due to small amount of Ti, and it is the first sign of order in the local structure of the solid solution. The ability of Ti atoms to move off the center of the oxygen octahedra and form well defined bonds results from the balance between electrostatic and short-range interactions, and favors ferroelectricity [42]. Local order around Ti, however, is not enough to explain the origin of the ferroelectric behavior; local polarizations must be correlated in order to reach a long-range polar order. The presence of Ti drives the net polarization, but the net polarization does not arise just from Ti-cells. As we observe in figure 5 (b) the three types of B-cell contribute in a similar way to the net polarization. Then, local order around Ti propagates in the structure favoring polar reorientation of Nb- and Mg-cells, and the linking between the ferroelectrically active cations is mediated by the host lattice. It was not possible to associate the macroscopic polarization with a well-defined region of the simulated sample. The results observed in this partially ordered configuration suggest that the crossover from a disorder state to an ordered one with *xPT* can be seen as percolation process where the correlation between local polarizations increases with the amount of Ti in the composition [50].

Two phase transitions are observed in the region as temperature increases (figure 6). The first one corresponds to a dipolar-glass transition, and it take place at the temperature at which there is a change in the slope of $q^{EA}$ (150 K for *x*=0.1 and 200 K for *x*=0.2). The change in slope occurs at the same place as a change in *W(p)* (inset figure 6). For temperatures below the transition, the asymmetric off-center peak in *W(p)* moves to lower polarizations with increasing temperature, and for temperatures above the transition there is an additional sharpening of the distribution. The higher temperature region is a glassy state with some dynamical polar degrees of freedom, which freezes into a ferroelectric state with decreasing temperature, consistent with experiment.[39] The second transition with increasing temperature corresponds to a transformation from a glassy state, to a true paraelectric with dynamical polar modes which average to zero, and is similar to the paraelectric phase observed in PMN.

*3.4. MPB*

The range of composition from x=0.3 to x=0.5 separates the region of rhombohedral symmetry from that of tetragonal symmetry. The presence of a compositional induced MPB in the simulated PMN-*x*PT is in agreement with the experimental observation although the calculated region is broader than the experimental one. In this transition zone various phases of different symmetries are present, which can be distinguished by cell distortions and polarization orientation. In this region, atomic displacements are more correlated, bond lengths narrow, and a higher degree of order in the structure is observable. For instance, the double peak at ~2 Å in the PDF for $x$=0.3 in figure 3 indicates that not only the Ti-O bond distances split, but also Nb-O and Mg-O. Signs of ordering are also observed in the splitting of peaks related with Pb: Pb-O and Pb-Nb/Mg/Ti. The order in the structure is translated to local polarizations. Instead of the broad distribution characteristic of the low Ti region, the components of the local polarizations are tightly distributed around mean values in this region (figure 3 (b)). The narrow distribution of each component of $W(p)$, which corresponds to a configuration with rhombohedral symmetry reveals that all local polarizations are aligned towards the average polar direction. Similar characteristics are present in the other configurations of the region, but the specific positions of the peaks depend on the particular symmetry of the considered structure.

Simulations show that the polar order caused by the presence of Ti overcomes the disorder produced by the distribution of Nb and Mg, and a ferroelectric state is obtained at low temperatures. Nevertheless, the degree of order is not strong enough to stabilize the structure along a particular direction. As a result, the region displays an intriguing interplay between long range and short-range effects. While the macroscopic polarization arises from a microscopically homogeneous configuration, the local polar direction depends on the particular B-cation assignment. We find that the MPB does not depend on the layout or B-cation distribution, and is only function of composition. Our results are compatible with the presence of an underlying flat free energy surface, which is a fundamental factor for polarization rotation mechanism [6]. As we observe in Table 2, the different phases present at a same composition are very close in energy. The small differences in B-cation assignment are responsible to stabilize one of the possible phases.

All configurations in this composition range become cubic at high temperatures. Nevertheless, the behavior below $T_C$ remains complex. In addition to the different structures observed at T = 0K, transitions depending on temperature and layout are also observed. Fluctuations in cell parameters and polarizations make it difficult to define unambiguously the different phases. Figure 7 shows the Cartesian components of $P$ as function of the temperature at $x$ = 0.3, 0.4 and 0.5. Each symbol in the figure describes a different layout while solid lines correspond to averages over configurations. We denote the component with small value as $P_x$, and the largest one as $P_z$. The results suggest that the main uncertainties at each temperature- composition point are mainly related to a particular component of $P$, which restricts $P$ to a particular plane.

At $x$=0.3, there are not big differences between layouts for the two larger components of $P$, but the behavior of the smallest one is unclear. $P_z \approx P_y$ and $P$ are confined to a plane between R and O phase. At low temperatures the phase is $M_B$ type. When temperature increases above 100 K, $P_x$ approaches zero, and the solid solution transforms to an O-phase at 300 K. Finally the three components of P become zero at $T_C$ = 400 K. Complex behavior is observed at $x$=0.4 where there is a change in the confinement plane of $P$ with temperature. $P_x$ is the easy component at low temperatures and the average symmetry is $M_B$ as in the previous composition. Nevertheless, the values of $P_y$ become unresolved for temperatures above 150 K. From this temperature, $P$ can easily move between O and T phase in an average $M_C$ phase. As temperature further increases, $P$ is approaching a T-phase before becoming cubic above $T_C$=500 K. Finally at $x$=0.5, the $M_C$ phase is observed for temperatures below 200 K while $P$ is definitively aligned along [00z] in the three layouts for temperatures above. This transition to a tetragonal phase also places the solid solution outside the MPB region, and shows that the right limit of the MPB shifts to lower concentrations with temperature. We note that the thinner MPB region at room temperature makes the simulated width closer to experiments. The system transforms to cubic at $T_C$=550 K.

We do not find any evidence in present study suggesting the presence of a $M_A$ (***P**=[xxz]*) phase as in the case of PZT[51]. Instead, we find an O phase present at the different concentrations and layouts, linking R and T phases. The intermediate O phase helps to connect the R and T phase in two steps. The first corresponds to a R-$M_B$-O sequence where the polarization can rotate in a (0-10) plane, and the second corresponds to an O-$M_C$-T sequence with a polar rotation in a (100) plane. Note that the low symmetry Tr phase is away from the probable path, and the presence may be due to numerical uncertain or size effects. Therefore, we estimate the probable polarization path across the MPB from the overall behavior of the region as R-$M_B$-O-$M_C$-T-C phase transformation sequence when increasing concentration and/or temperature. The possibility of such phase sequence in ferroelectric perovskites was theoretically predicted with an eight-order expansion of the Devonshire theory [52], and it was found in different experimental studies in PMN-*x*PT[53] and PZN-*x*PT[54].

Our model simulations are able to reproduce the MPB region experimentally observed in PMN-*x*PT. We show that it is possible to describe such behavior in an ideal system and without the presence of a complex microstructure like polar domains or defects [5]. We find that the MPB is induced only by composition while B-cation distribution affects the relative stability between the different phases observed in the region.

*3.5. High Ti*

The larger amount of Ti in the region is enough to stabilize the tetragonal structure, and the solid solution displays a behavior similar to PT for $x \geq 0.6$. Nevertheless, structural and polar disorder remains until Mg and Nb are completely substituted by Ti. Bond distances are better defined, and the PDF displays sharper peaks as it is shown for *x*=0.9 in figure 3 (a). In this case, for instance, the first B-O bond splits in four different peaks, where each one corresponds mainly to Ti-O, Nb/Mg-O, Ti/Nb-O and Mg-O respectively. Local polarizations are also more concentrated, and *W(p)* has sharper peaks at the macroscopic polar value in figure 3 (b). In this case, small peaks also emerge along both polar and non-polar directions reflecting the presence of Nb and Mg in the composition. In fact, these peaks describe the effects that take place when isolated Nb and Mg are introduced in pure PT. Such effects induce the reduction of the tetragonal distortion, spontaneous polarization and the increment of volume. When a Nb atom substitutes a Ti in PT, it experiments a larger displacement along polar direction from the center of oxygen octahedral due to its bigger charge while the stronger Pb-Nb repulsion produces the increment in the cell volume and the reduction of *c/a*. As a result, the local polarization of the cell reduces from 66.7 $\mu C/cm^2$ to 57.3 $\mu C/cm^2$. Neighboring cells are little affected. Whereas the upper and lower cells diminish slightly their values, those that are to the sides display a small component (2$\mu C/cm^2$) against the impurity. When an Mg replaces a Ti, on the other side, it experiences a smaller off-center displacement along polar axis due to its more ionic character, and its cell polarization slightly decreases to 63.7 $\mu C/cm^2$. The changes in volume and in *c/a* are similar to the ones produced by the Nb but less pronounced. Nevertheless, the stronger Mg-O repulsion expands oxygen octahedra and produces significant changes in neighboring cells. The polarization of the upper cell is reduced to 52.4 $\mu C/cm^2$, while lateral cells present a component normal to the polar axis and pointing towards the Mg of 6.5 $\mu C/cm^2$. Then, that the small peaks observed in *W(p)* indicate that Nb and Mg atoms act as isolated impurities at this concentration.

Figure 8 shows the temperature evolution of tetragonal distortion at various compositions in the high Ti content region. At all composition *c/a* decreases with temperature, and there is just one phase transition to the high temperature cubic phase. The Curie temperature increases linearly with *x*, from 650 K at *x*=0.6 to 850 K for *x* =1.0. In this case, the value obtained in the simulations is in reasonable agreement the experimental one of 763 K, although the model overestimation of $T_C$ could be related to the larger observed tetragonal distortion.

## 4. Discussions and Conclusions

The results above described were used to build the temperature-concentration phase diagram presented in figure 9. As a reference, the figure also displays an accepted experimental one [55]. Several phase diagram of PMN-$x$PT have been proposed in the last years [2,55,56]. They are essentially similar, and the differences between them are limited mainly to the stability of different phase in MPB region or close to it. While the precise details in the description are fundamental for a deeper understanding of the phase diagram, they are not relevant for present study. The precision in our study is limited by size effects, the interval of concentration, temperature step, and the particular choice of the B-site cation distribution. In addition, we consider only transitions that involve the polarization as order parameters. Taking into account such limitations, the simulated phase diagram is in excellent qualitative and quantitative agreement with the experiments. The theoretical phase diagram correctly reproduces the four regions observed in experiments as temperature and compositions are varied. At high temperature, the solid solution is cubic (C) at all composition with a Curie temperature that increases essentially linearly with $x$. Below $T_C$ the left side is denoted as R, the right side is tetragonal (T), and the region in between corresponds to the MPB. This region emerges naturally with composition and displays a behavior which is consistent with the easy rotation of the polarization direction. The simulated region starts at a composition in agreement the experiments, but it extends to compositions greater than those observed experimentally. The size overestimation of the region could indicate limitations in the description or it could be related to size effects. Since behavior is very sensitive to the configurations, simulations with larger cells may help to narrow the region.

According to our simulation, the behavior of the solid solution is due to presence of different atoms at the equivalent B-sites. Whereas Mg and Nb favor a disordered state, Ti favors an ordered one. The degree of order gradually changes with composition, and hence the solid solution behaves as a complex system where both characteristics coexist. At $x = 0$, local polarizations are randomly distributed, and PMN behaves as a polar glass. The addition of Ti favors the correlation between cells, and a weak polar order arises in the low Ti region. More cells are reoriented with the increment of Ti, and polar order overcomes disorder in the MPB region. Every single cell prefers to align its polarization towards the average direction, and the ferroelectric state is clearly developed. Nevertheless the level of order in the region is not enough to stabilize the tetragonal structure, and a delicate stability between the different phases is observed. Finally, polar order completely dominates the system behavior at even higher Ti contents, and solid solution behaves similar to PT in the high Ti region.

The second aspect corresponds to the representation of the atomic order. Compositional order can greatly influence the properties of the solid solution, and the adopted distribution was able to capture essential features of the real system. In particular, our results indicate that the intrinsic degree of randomness in the RSM distribution is enough to frustrate structural and long-range polar order, and to describe the relaxor features of PMN. Contrary to what is believed, we show that behavior of PMN can be explained without the presence of PNRs. Recent MD simulations were able to describe the dielectric response of the relaxor PMN-0.25PT without PNRs [17], so both results raise the question about the precise role played by PNRs in the relaxor behavior

Acknowledgments

This work was sponsored by Consejo Nacional de Investigaciones Científicas y Tecnológicas de la República Argentina (CONICET) and the US Office of Naval Research grant number N00014-07-1-0451. We thank Q. Peng for helpful discussions.

**Table 1.** Shell model parameters for PMN and PT. Units of energy, length and charge are given in eV, Å, and electrons.

| Atom | Core charge | Shell charge | $k_2$ | $k_4$ |
|------|-------------|--------------|-------|-------|
| Nb   | 5.1464      | -0.3506      | 75.35 | 26896.6 |
| Mg   | 2.4628      | -0.1046      | 85.75 | 0.0 |
| Nb   | 5.5279      | -2.3798      | 713.64 | 3983.9 |
| Ti   | 9.7297      | -6.8449      | 1937.88 | 0.0 |
| O    | 0.7057      | -2.2659      | 26.48 | 1324.4 |

| Short range | A | B | $\rho$ |
|-------------|---|---|--------|
| Pb-O  | 6291.34 | 296.28  | 0.265259 |
| Mg-O  | 1042.14 | 63.48   | 0.315781 |
| Nb-O  | 1508.13 | 3.78    | 0.298781 |
| Ti-O  | 1416.39 | 3.68    | 0.290791 |
| O-O   | 283.41  | -103.27 | 0.520557 |

**Table 2**. Structural characteristics obtained with the model for PMN-$x$PT in the MPB region. Energy differences are in meV/atom and polarizations are in $\mu C/cm^2$. Note that the symmetries of the phases are approximate.

| | | $x = 0.3$ | | | |
|---|---|---|---|---|---|
| Layout | Phase | $\Delta E$ | $P_x$ | $P_y$ | $P_z$ |
| 1 | R | 0.00 | 22 | 23 | 24 |
| 2 | $M_B$ | +0.50 | 18 | 24 | 24 |
| 3 | O | +2.43 | 1 | 28 | 28 |

| | | $x = 0.4$ | | | |
|---|---|---|---|---|---|
| Layout | Phase | $\Delta E$ | $P_x$ | $P_y$ | $P_z$ |
| 1 | Tr | 0.00 | 20.2 | 24.5 | 29.6 |
| 2 | O | -0.04 | 0.1 | 30.6 | 30.4 |
| 3 | O | +0.70 | 1.7 | 30.2 | 30.3 |

| | | $x = 0.5$ | | | |
|---|---|---|---|---|---|
| Layout | Phase | $\Delta E$ | $P_x$ | $P_y$ | $P_z$ |
| 1 | O | 0.00 | 3.3 | 30.9 | 33.0 |
| 2 | $M_C$ | +0.02 | 1.3 | 28.0 | 35.6 |
| 3 | T | +0.36 | 0.7 | 2.8 | 42.9 |

Figure captions

**Figure** : Volume of PMN-*x*PT relaxed structures as function of Ti content obtained from the simulations. For comparison experimental values extrapolated to 0 K are included at *x*=0 [25], 0.1[6], 0.2 [26], 0.3 [5], 0.39 [5], 0.5 [12], 0.65[12] and 1[27], and LDA results from PMN and PT.

**Figure 2:** (Color online): Lattice parameters (a), and magnitude and components of the polarization (b) as function of Ti-content in PMN-*x*PT relaxed structures obtained from the simulations. The square, circle and triangle symbols represent different layouts respectively and the solid lines link averages over configurations. The magnitude of the polarization $P_s$ does not show a significant dependence in the layouts.

**Figure 3**: Pair distribution function, PDF, (a) and polar distribution function components, $W(p)$, (b) obtained from the simulations in relaxed structures of 8640-atom supercells of PMN-*x*PT at *x* = 0, 0.2, 0.3, and 0.9. The better defined bond distances that take place with the increase in *x* produce sharper distributions of local polarizations.

**Figure 4**: Polar distribution function along a pseudo cubic component for PMN at different temperatures obtained from the simulations. Due to the cubic symmetry the three components of $W(p)$ are equivalent. The inset shows the corresponding temperature evolution of the Edwards-Anderson order parameter.

**Figure 5**:Individual Ti-O, Mg-O and Nb-O bond length distributionsfor the shortest B-O pair (a) and individual cell-type distributions of local polarizations (b) at *x*=0.2. In this low Ti region glassy and ferroelectric characteristics coexist.

**Figure 6:** (Color online): Temperature evolution of a component of the Edwards-Anderson order parameter, $q^{EA}$, in PMN-*x*PT for *x*=0.1 and *x*=0.2. The inset shows $W(p)$ for *x*=0.2 at different temperatures. At these compositions, there are no significant differences between layouts.

**Figure 7:** (Color online): Average polarization components of PMN-*x*PT in the MPB region as function of the temperature at *x* 0.3 (a), 0.4 (b), and 0.5(c). The square, circle and triangle symbols represent different layouts respectively and the solid lines link averages over configurations. Despite layouts have the same type of order; results in this region are sensitive to B-cation assignment.

**Figure 8**: Tetragonal distortion as function of the temperature in the high Ti region of PMN-*x*PT. In this region, all layouts display similar results.

**Figure 9:** (Color online): Temperature-Composition phase diagram of PMN-$x$PT estimated from model simulations (red) and by experiments (blue)[55]. The represented regions are cubic (C), rhombohedral (R), morphotropic phase boundary (MPB), and tetragonal (T). Model simulations correctly reproduce the successive phase transformations experimentally observed with composition and temperature. The points show T-X conditions for MD simulations we performed.

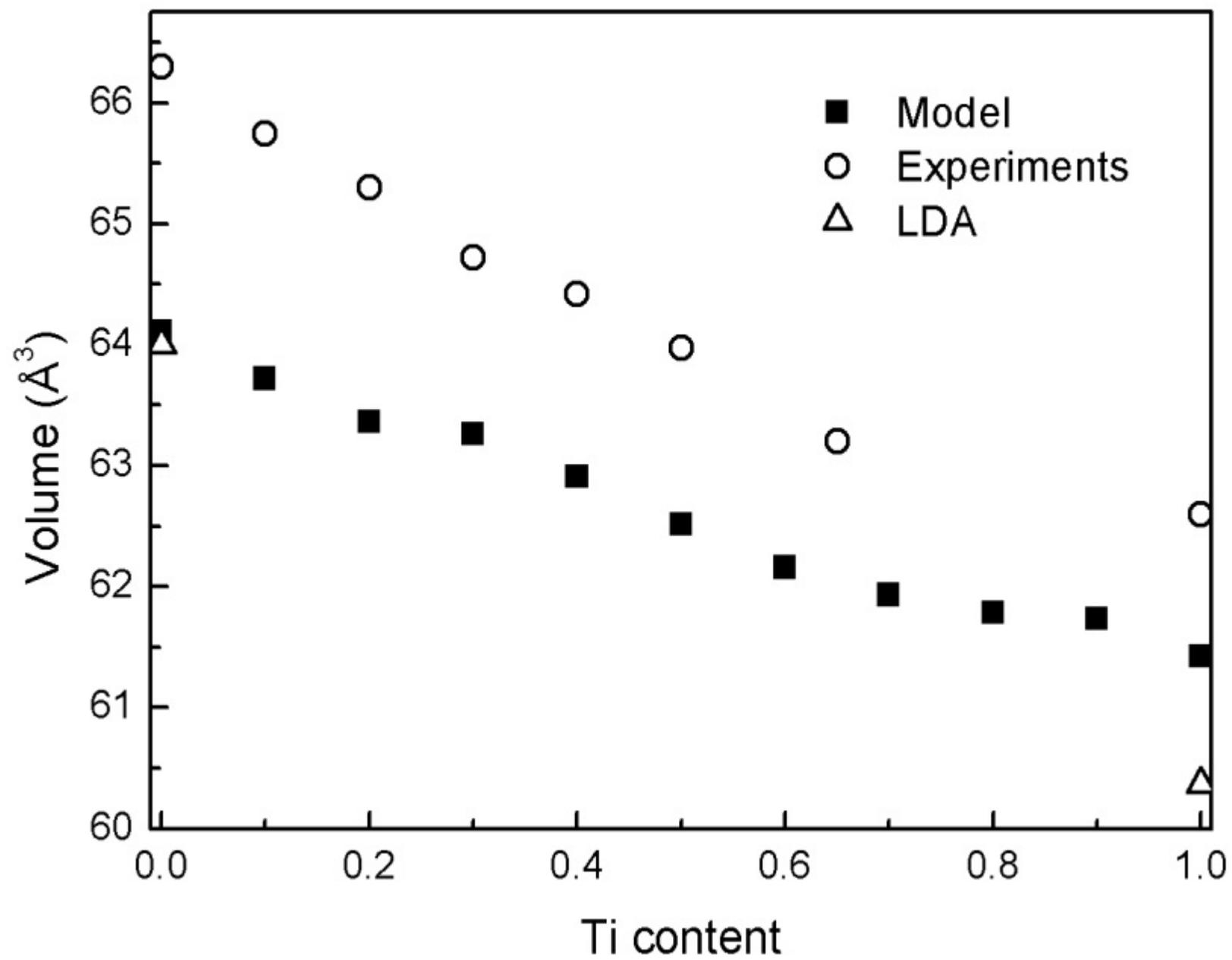

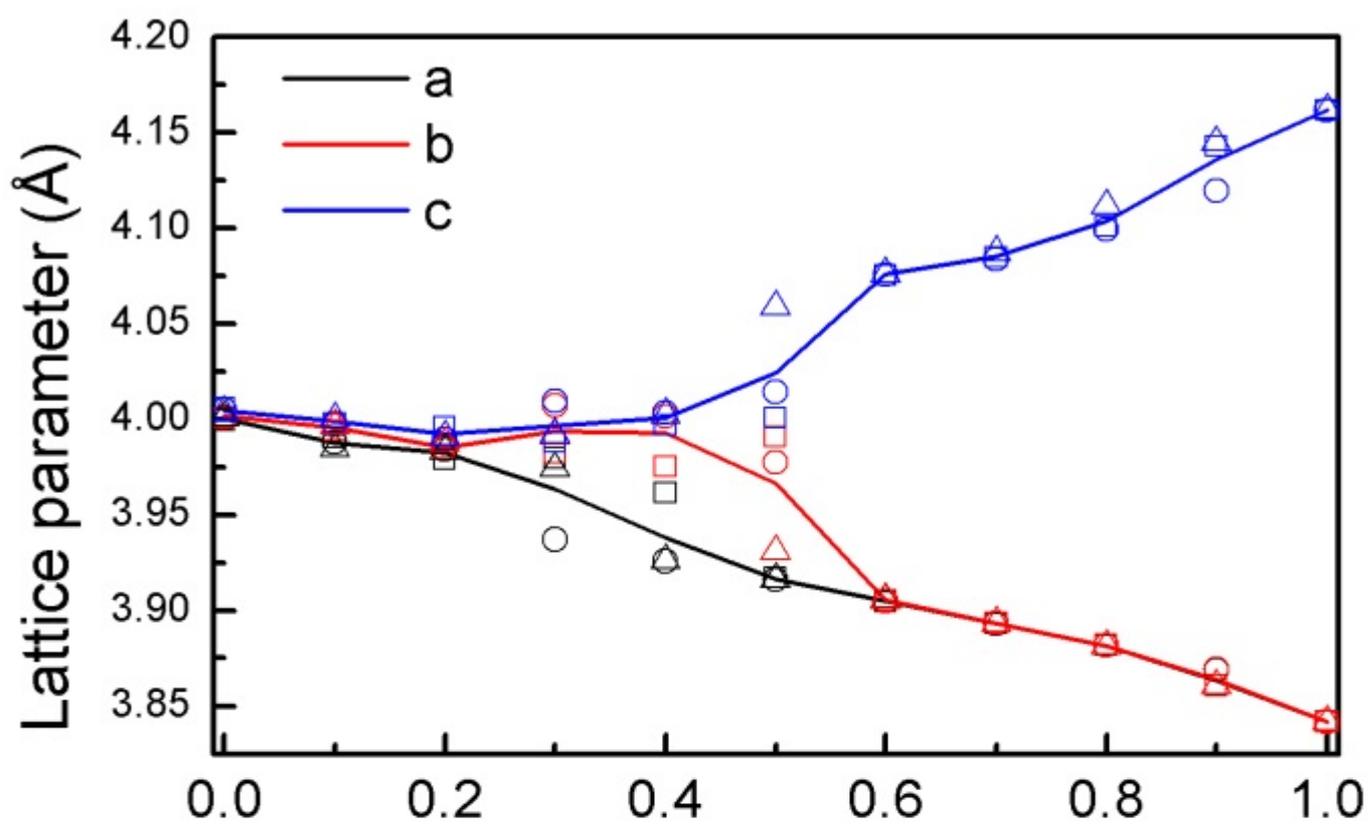
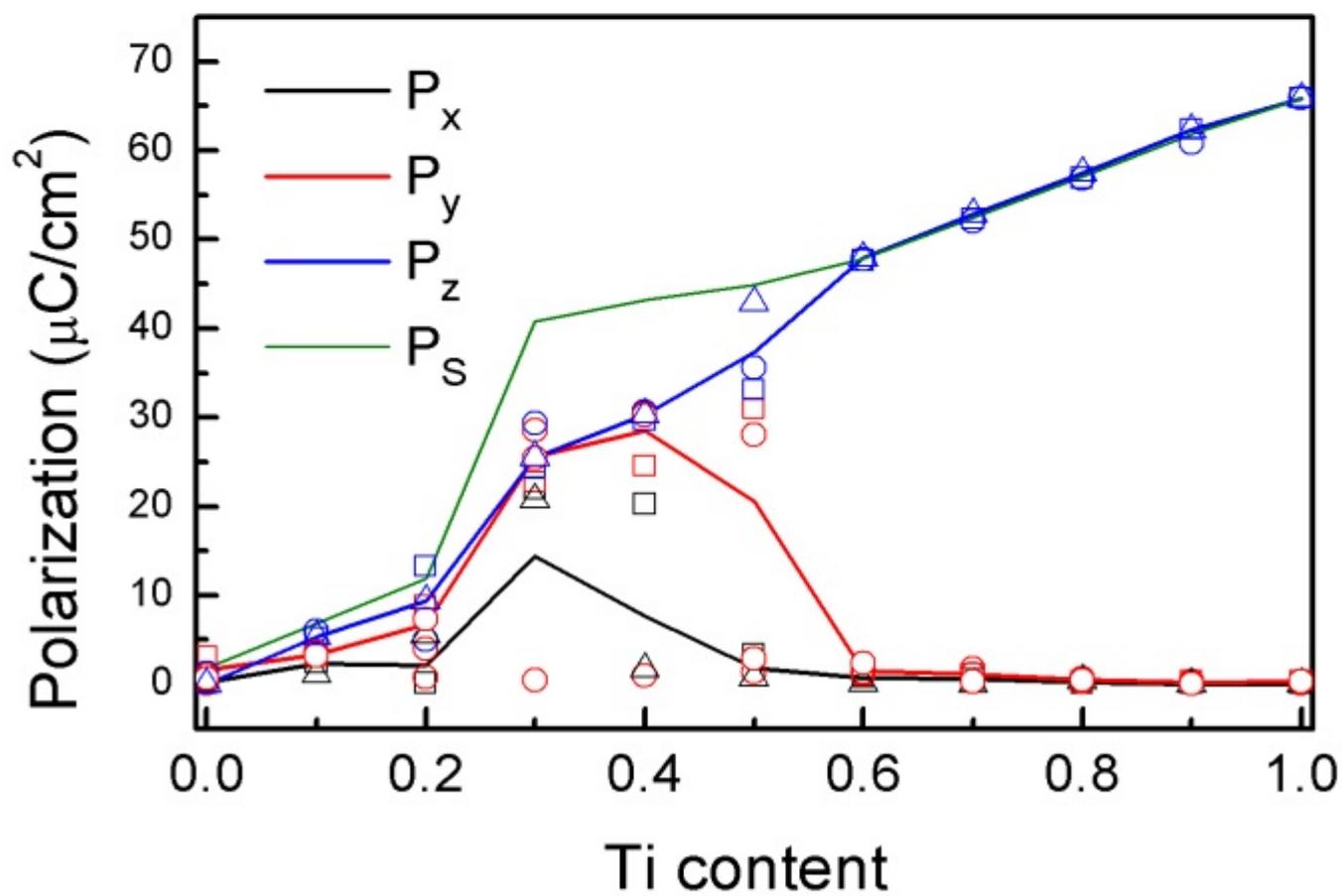

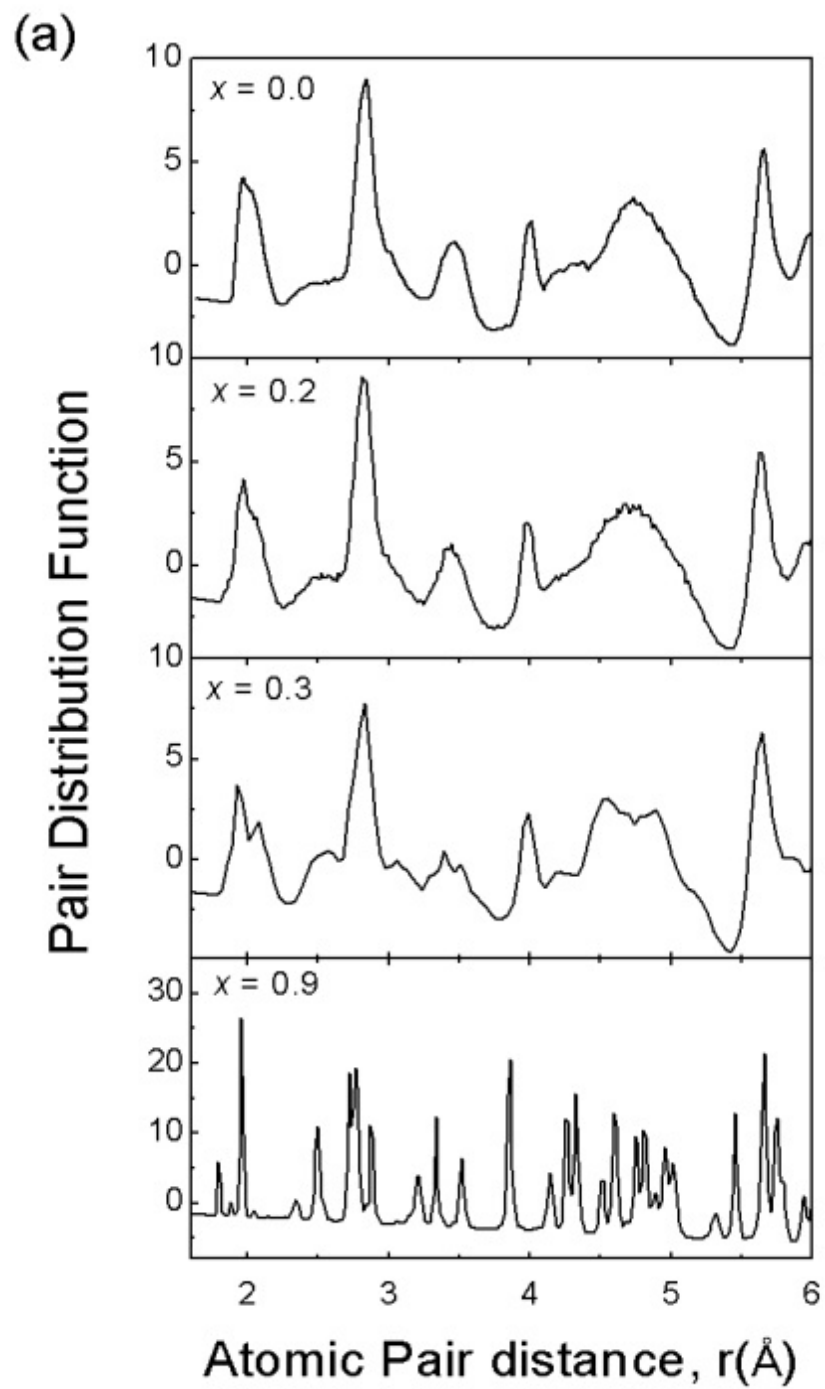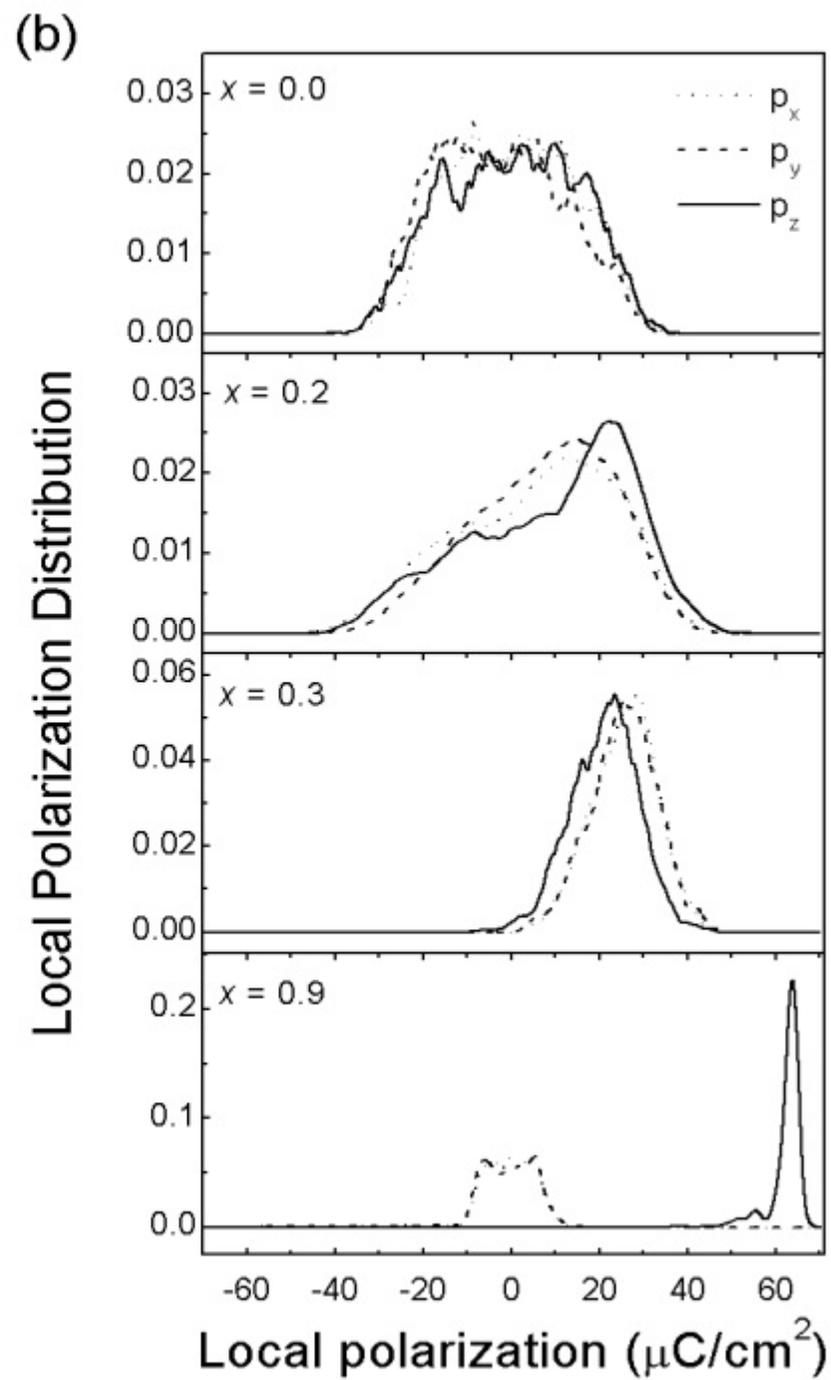

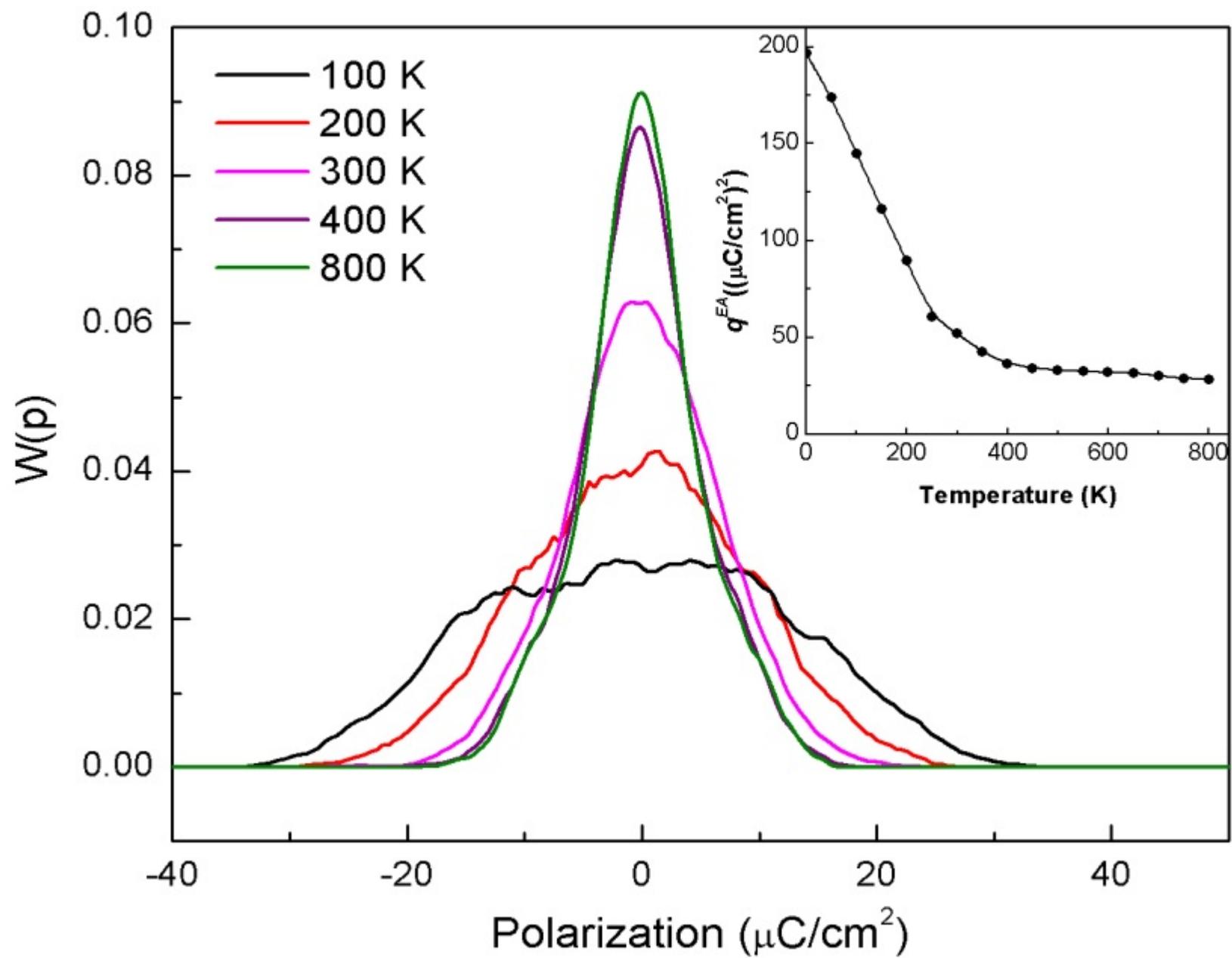

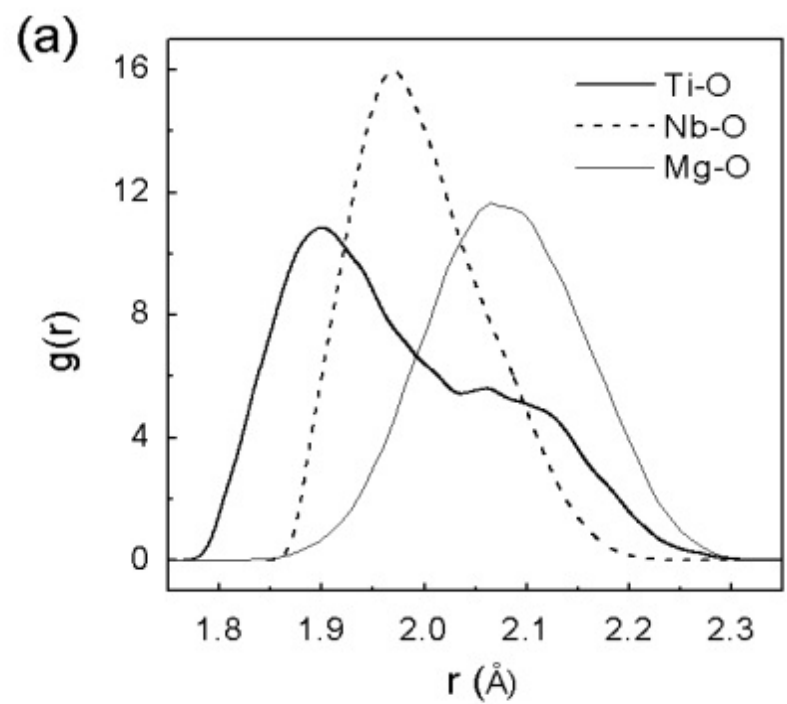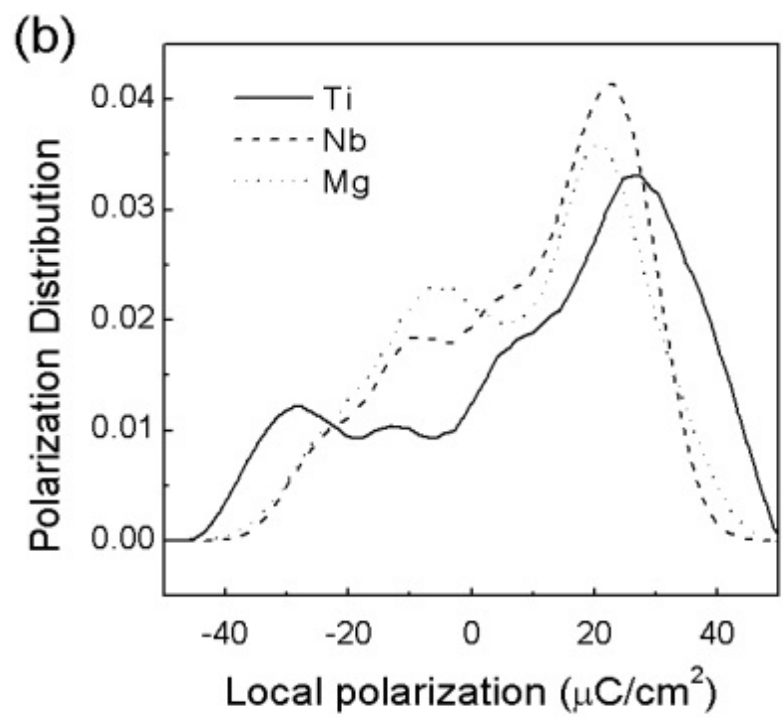

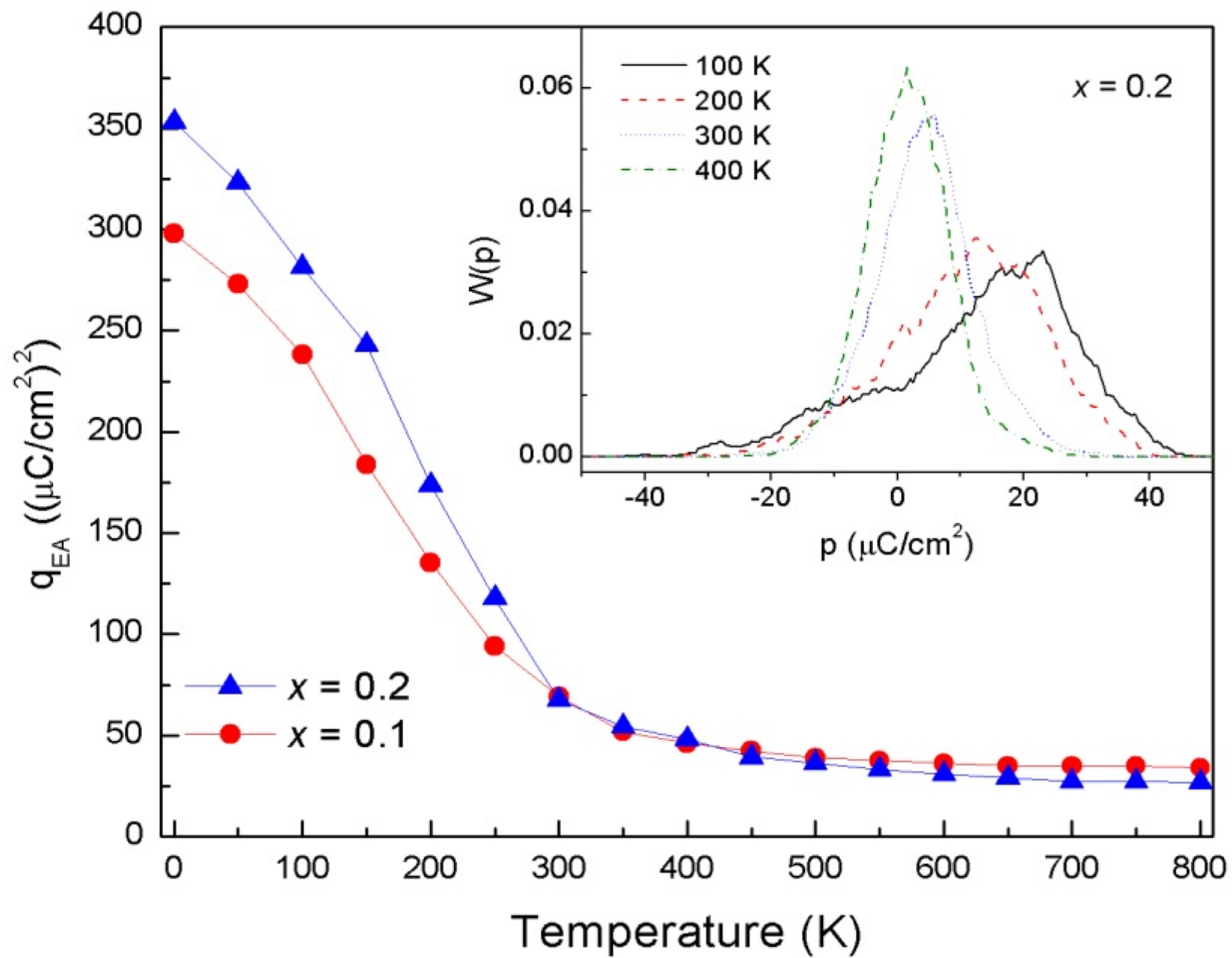

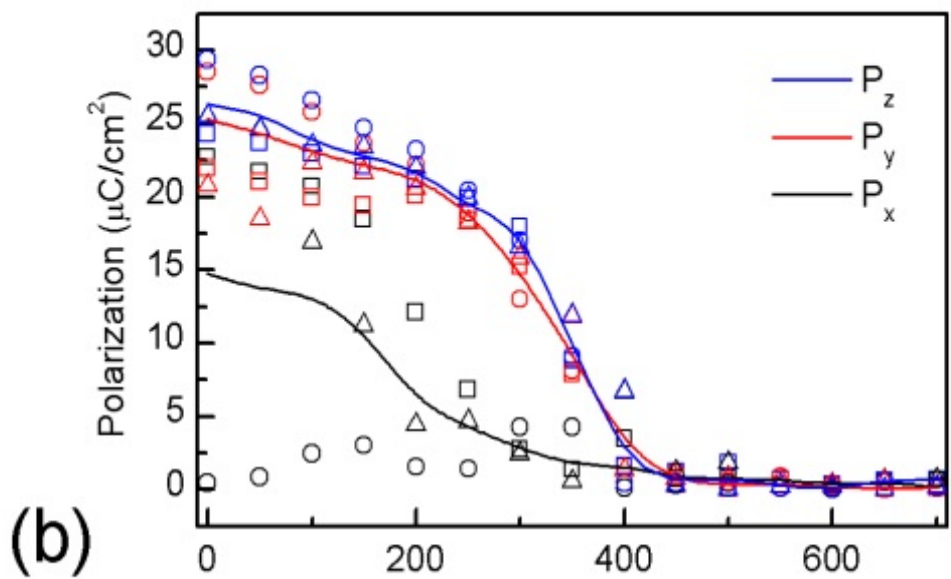

(b)

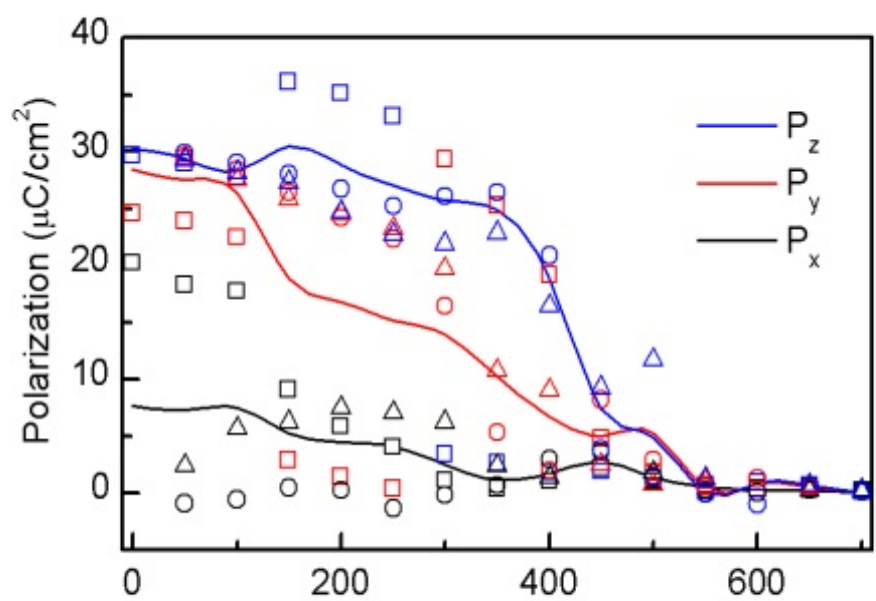

(c)

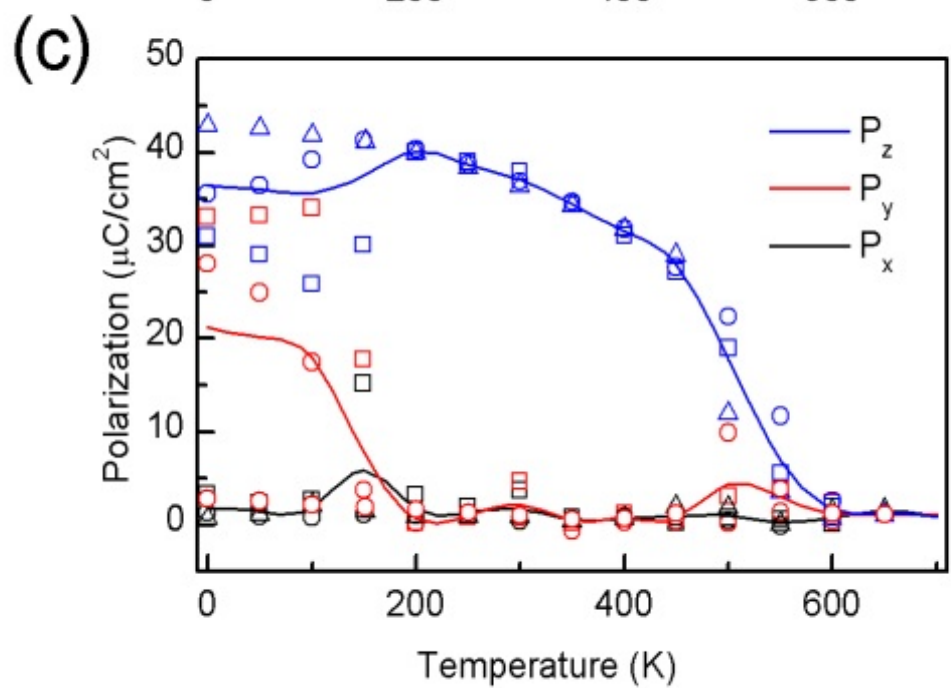

Temperature (K)

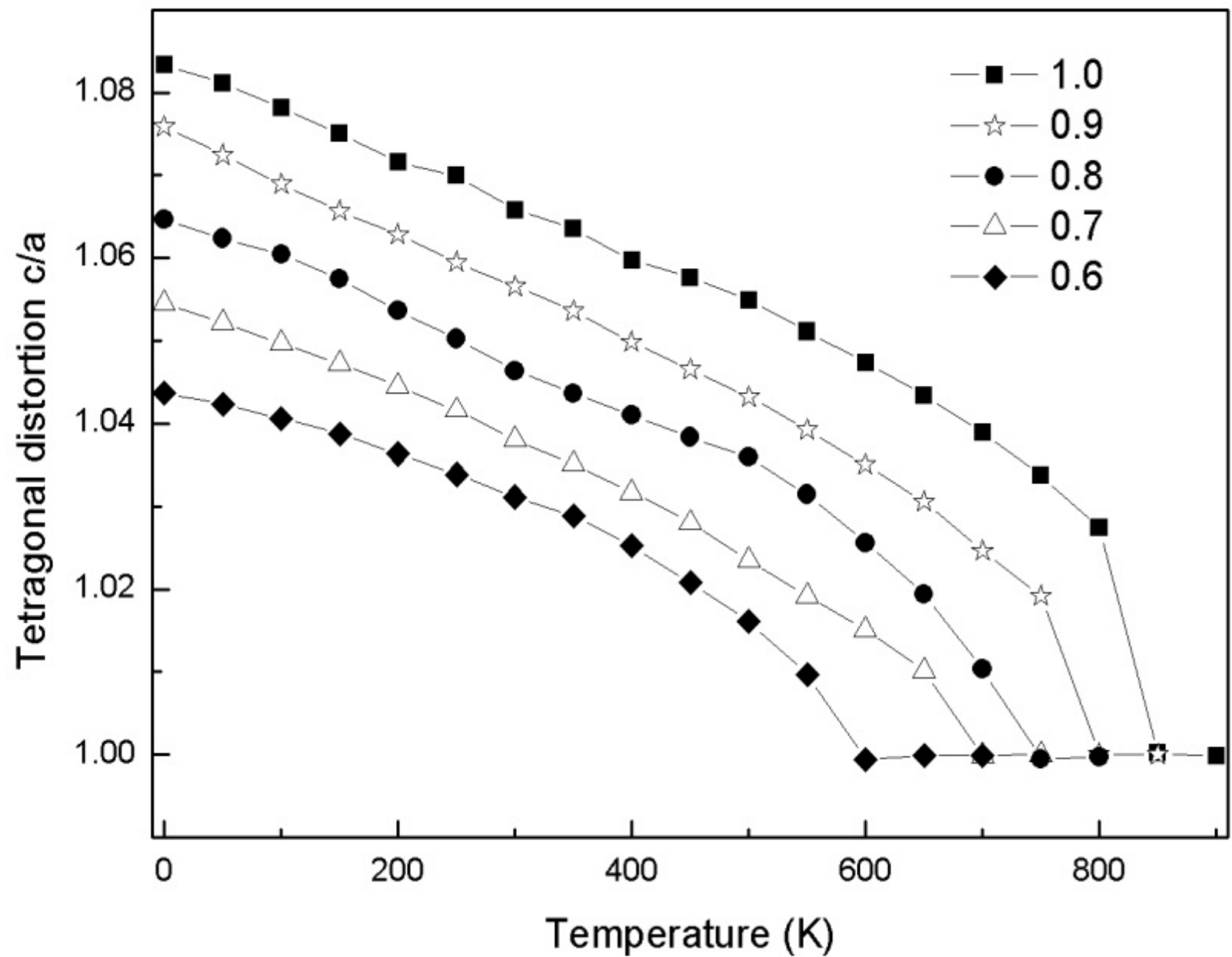